\pdfoutput=1
\documentclass[11pt]{article}
\usepackage{amssymb}
\parskip \baselineskip

\newcommand{\ra}{\rightarrow}

\begin{document}

\title{ALLSAT compressed with wildcards: Frequent Set Mining}

\author{Marcel Wild}

\maketitle

\section{Introduction}

This is the first draft of a hopefully longer article in spe for the series 'ALLSAT compressed with wildcards'. 
 The author never published on Frequent Set Mining  before (but knows related data mining frameworks such as Formal Concept Analysis, Knowledge Spaces and Relational Databases). He therefore hopes that this draft attracts co-authors  helping to pit his algorithms against state-of-the-art methods such\footnote{These three compete against {\it each other} in [H].} as Apriori, Eclat, $FP$-growth to which he has no access.  The two newcomer algorithms are called  Find-All-Facets and Facets-To-Faces.

In a nutshell Frequent Set Mining, aka FSM, goes like this (more fleshed out introductions are easy to google):  Consider an arbitrary binary table whose columns are labelled by {\it items} (such as bread, butter etc. sold in a supermarket) and whose rows are called {\it transactions} (matching the itemsets bought by customers during a specific day). Fix any natural number $\alpha$, called the {\it threshold}, and call an itemset $X$ {\it frequent} if $X$ is a subset in at least $\alpha$ many transactions. Stripped to its core FSM attempts to display all frequent sets in a succinct way. The family of all frequent sets constitutes a simplicial complex  ${\cal S}{\cal C}$ and our first algorithm finds its maximal members (=facets). The second algorithm uses the facets to compress the {\it whole} of  ${\cal S}{\cal C}$.

Here comes the Section break-up.  In Section 2 our first toy database (Table 1) allows to find the four maximal frequent sets by inspection. Feeding them to the algorithm Facets-To-Faces from [W2]  reveals that there are 14911 frequent sets altogether and they can be densely packed by the use of wildcards. While the next two Sections prepare the last Section they are interesting in their own right. Section 3 reviews and refines a technique called 'Vertical Layout' that speeds up finding the maximal sets within arbitrary set families. Section 4 presents a novel method to find the facets of  any a priori unknown (but 'decidable') simplicial complex ${\cal S}{\cal C}$. This Find-All-Facets algorithm replaces the expensive Dualize+Advance subroutine (or its variants) by  multiple applications of Vertical Layout.
 Section 5 walks the reader through Find-All-Facets when ${\cal S}{\cal C}$ is the family of all frequent sets of a second  toy database (Table 4). Afterwards, as in Section 2, the seven found facets are handed over to Facets-To-Faces to compress the whole of ${\cal S}{\cal C}$ .  

All sets in this article are assumed to be {\it finite}.

\section{The first toy database}

 For convenience we take  $U=[w]:=\{1,2,\ldots, w\}$ as our set of items.
In our first binary Table 1 (which for better visualization uses x and blanks instead of 1 and 0) the maximal frequent sets (=facets) are easily determined. Specifically, if $\alpha:=2$ then $F_1:=[16]\setminus\{1,2,3,4\}$ is a frequent itemset because it is contained in the transactions $T_1$ and $T_2$. Obviously $F_1$ is maximal. Likewise 
$F_2:=[16]\setminus\{5,6,7,8\}$ and $F_3:=[16]\setminus\{9,10,11,12\}$ and $F_4:=[16]\setminus\{13,14,15,16\}$ are facets. We leave it as an exercise to verify that $F_1$ to $F_4$ are the {\it only} maximal facets.

\begin{tabular}{c|c|c|c|c|c|c|c|c|c| c|c|c|c|c|c|c| }
& 1 & 2 & 3& 4 & 5 & 6 &7 & 8 & 9  & 10 & 11 & 12& 13 & 14 & 15 &16 \\ \hline 
&  &  & &  &  &  & &  &   &  &  & &  &  &  & \\ \hline
$T_1=$ & & & & & $x$ & $x$ & $x$ & $x$ & $x$ & $x$ & $x$ & $x$ & $x$ & $x$ & $x$ & $x$  \\ \hline
$T_2=$ & & & & & $x$ & $x$ & $x$ & $x$ & $x$ & $x$ & $x$ & $x$ & $x$ & $x$ & $x$ & $x$  \\ \hline

$T_3=$ &  $x$ & $x$ & $x$ & $x$ & & & & & $x$ & $x$ & $x$ & $x$ & $x$ & $x$ & $x$ & $x$  \\ \hline
$T_4=$ &  $x$ & $x$ & $x$ & $x$ & & & & & $x$ & $x$ & $x$ & $x$ & $x$ & $x$ & $x$ & $x$  \\ \hline

$T_5=$ &  $x$ & $x$ & $x$ & $x$ & $x$ & $x$ & $x$ & $x$ & & & & & $x$ & $x$ & $x$ & $x$  \\ \hline
$T_6=$ &  $x$ & $x$ & $x$ & $x$ & $x$ & $x$ & $x$ & $x$ & & & & & $x$ & $x$ & $x$ & $x$  \\ \hline

$T_7=$ &  $x$ & $x$ & $x$ & $x$ & $x$ & $x$ & $x$ & $x$ &  $x$ & $x$ & $x$ & $x$ & & & & \\ \hline
$T_8=$ &  $x$ & $x$ & $x$ & $x$ & $x$ & $x$ & $x$ & $x$ &  $x$ & $x$ & $x$ & $x$ & & & & \\ \hline
 \end{tabular}

{\sl Table 1: The four maximal frequent sets of this database are found by inspection}

Hence the simplicial complex ${\cal F}{\cal S}_1$ of all frequent sets is ${\cal F}{\cal S}_1={\cal P}(F_1)\cup\cdots\cup {\cal P}(F_4)$. Unfortunately this union of powersets is not disjoint; for instance $\{1,2,3,4\}$ belongs to three powersets. We can make the union disjoint (indicated by $\uplus$) as follows:

(1)\quad
${\cal P}(F_1)\uplus\bigg({\cal P}(F_2)\setminus{\cal P}(F_1)\bigg)\uplus\bigg({\cal P}(F_3)\setminus\Big({\cal P}(F_1)\cup{\cal P}(F_2)\Big)\bigg)
\uplus\bigg({\cal P}(F_4)\setminus\Big({\cal P}(F_1)\cup{\cal P}(F_2)\cup{\cal P}(F_3)\Big)\bigg)$

This is achieved neatly by applying the Facets-To-Faces algorithm of [W2] to the facets $F_1$ to $F_4$:

\begin{tabular}{c|c|c|c|c|c|c|c|c|c| c|c|c|c|c|c|c|c }
& 1 & 2 & 3& 4 & 5 & 6 &7 & 8 & 9  & 10 & 11 & 12& 13 & 14 & 15 &16 \\ \hline 
&  &  & &  &  &  & &  &   &  &  & &  &  &  & \\ \hline
$r_1=$ &{\bf 0} & {\bf 0} &{\bf 0} &{\bf 0} & $2$ & $2$ & $2$ & $2$ & $2$ & $2$ & $2$ & $2$ & $2$ & $2$ & $2$ & $2$ & $4096$  \\ \hline

$r_2=$ &  $e$ & $e$ & $e$ & $e$ &{\bf 0} & {\bf 0}&{\bf 0} &{\bf 0} & $2$ & $2$ & $2$ & $2$ & $2$ & $2$ & $2$ & $2$ & $3840$ \\ \hline

$r_3=$ &  $e_1$ & $e_1$ & $e_1$ & $e_1$ & $e_2$ & $e_2$ & $e_2$ & $e_2$ &{\bf 0} & {\bf 0}&{\bf 0} &{\bf 0} & $2$ & $2$ & $2$ & $2$ & $3600$ \\ \hline

$r_4=$ &  $e_1$ & $e_1$ & $e_1$ & $e_1$ & $e_2$ & $e_2$ & $e_2$ & $e_2$ &$e_3$ &$e_3$ &$e_3$ &$e_3$ &{\bf 0} & {\bf 0} &{\bf 0}&{\bf 0} & $3375$  \\ \hline
\end{tabular}

{\sl Table 2: Compressed representation of ${\cal F}{\cal S}_1$}

Specifically, we use the don't-care symbol '2' to indicate that a bit at this position is free to be 0 or 1. Hence the row $r_1$ comprises $2^{12}= 4096$ bitstrings and matches the powerset ${\cal P}(F_1)$. More subtle, the wildcard $ee...e$ means 'at least one $1$ here'. It follows that 
$r_2={\cal P}(F_2)\setminus{\cal P}(F_1)$ contains $(2^4-1)\cdot 2^8$=3840 bitstrings. Similarly $r_3,\ r_4$ are compact ways to write the second and third set difference appearing in (1). One concludes that the `database' given by Table 1 has  $4096+3840+3600+3375 = 14911$ frequent sets.

{\bf 2.1} The format of Table 2 invites a finer statistical analysis of ${\cal F}{\cal S}_1$. Let us sketch three alleys. 

 First, by [W2, Section 5], for any fixed $k$ the number of $k$-element bitstrings within a $012e$-row is readily calculated as the coefficient at $x^k$ of some suitable polynomial. For instance, a moment's thought shows that the polynomial for $r_3$ is $((1+x)^4-1)^2 (1+x)^4$. Since the coefficient at $x^7$ is 776, there live 776 frequent sets of cardinality 7 in $r_3$. Adding three similary obtained numbers to 776 one finds that altogether there are exactly  3120 frequent sets of cardinality 7.

Second, the number of frequent sets containing any fixed set $X$ is easy to obtain. To witness, if $X = \{7, 8,9\}$ then  this number is
$$|r_1 \cap X | + |r_2 \cap X| + |r_3 \cap X| + |r_4 \cap X| = 512 + 0 + 0 + 480 = 992.$$

Third, by running Facets-To-Faces twice one can count how many among the $\alpha$-frequent sets are not $(\alpha+1)$-frequent.

\section{Using Vertical Layout to find all maximal members in arbitrary set families}

So called {\it Vertical Layout} is common  in the FSM-literature, probably first introduced  in [HKMT, p.151].  In the same year, i.e. 1995, the underlying idea was independently discovered in [W1,p.113]. Vertical Layout (VL) can  be used to determine the family $Max({\cal G})$
of all (inclusion-) maximal sets in {\it any} set family ${\cal G}$. It is most efficient if ${\cal G}$ consists of many small sets as opposed to few large sets. Let us illustrate the VL technique on the set family
${\cal G}_0:=\{X_1,\ldots, X_{15}\}$ defined by Table 3. 

\begin{tabular}{l|c|c|c|c|c|c|c|c|c|c} 
   index & 1 & 2& 3 & 4 & 5 &6 & 7 & 8 & 9 \\ \hline
$1$ &   $0$   &   $1$ &  $0$  &  $0$  &  $1$  &  $0$   &   $1$  &   $0$  &   $0$& \\    \hline
$2$ &   $1$   &   $1$ &  $0$  &  $0$  &  $0$  &  $1$   &   $1$  &   $0$  &   $0$& \\ \hline
$3$ &   $0$   &   $1$ &  $0$  &  $1$  &  $1$  &  $0$   &   $1$  &   $1$  &   $1$& \\ \hline
$4$ &   $0$   &   $0$ &  $1$  &  $1$  &  $0$  &  $0$   &   $0$  &   $0$  &   $1$& \\ \hline
$5$ &   $0$   &   $1$ &  $1$  &  $0$  &  $1$  &  $0$   &   $0$  &   $1$  &   $0$& \\ \hline
$6$ &   $0$   &   $1$ &  $1$  &  $0$  &  $0$  &  $0$   &   $0$  &   $1$  &   $0$& \\ \hline
$7$ &   $1$   &   $1$ &  $0$  &  $0$  &  $1$  &  $1$   &   $1$  &   $0$  &   $1$& \\ \hline
$8$ &   $1$   &   $0$ &  $1$  &  $1$  &  $0$  &  $0$   &   $1$  &   $0$  &   $0$& \\ \hline
$9$ &   $0$   &   $0$ &  $0$  &  $1$  &  $0$  &  $0$   &   $0$  &   $0$  &   $0$& \\ \hline
$10$ &   $0$   &   $1$ &  $1$  &  $0$  &  $1$  &  $0$   &   $0$  &   $1$  &   $0$& \\ \hline
$11$ &   $0$   &   $0$ &  $0$  &  $0$  &  $1$  &  $0$   &   $0$  &   $1$  &   $0$& \\ \hline
$12$ &   $0$   &   $1$ &  $0$  &  $0$  &  $0$  &  $0$   &   $0$  &   $1$  &   $1$& \\ \hline
$13$ &   $0$   &   $0$ &  $0$  &  $0$  &  $0$  &  $1$   &   $1$  &   $1$  &   $0$& \\ \hline
$14$ &   $1$   &   $0$ &  $1$  &  $1$  &  $0$  &  $1$   &   $1$  &   $0$  &   $1$& \\ \hline
$15$ &   $0$   &   $1$ &  $0$  &  $0$  &  $1$  &  $0$   &   $0$  &   $0$  &   $0$& \\ \hline
\end{tabular}

{\sl Table 3: Illustrating Vertical Layout}

Crucial is the 'vertical view' of Table 3, i.e. we record the locations of the $0$'s in the $i$-th column:

$Zeros[i]:=\{j\in [15]:\ j\not\in X_i\}\hspace{1cm}  (1\le i\le 9)$

Furthermore put

$Card[i]:=\{j\in [15]:\ |X_j|=i\}\hspace{1cm} (1\le i\le 6) $

$AllMaxsets:=\{ j\in [15]:\ X_j\in Max({\cal G}_0\}) \hspace{1.2cm}\hbox{\rm (unknown) }$

These sets stay fixed. In contrast $FoundMaxsets$, which contains the  indices found so far, keeps on growing. Throughout the algorithm we will have

$FoundMaxsets\subseteq AllMaxsets\subseteq Candidates$,

where $Candidates$ is a certain shrinking set of indices. The algorithm terminates as soon as $FoundMaxsets=Candidates$.
 Initially we put $FoundMaxsets:=\emptyset$ and $Candidates:=[15]$. 
The sets with indices  in $Card[6]=\{3,7,14\}$ have maximum cardinality and thus are\footnote{We could thus put $FoundMaxsets:=Card[6]$. But for systematic reasons (concerning duplicated sets) we postpone this assignment.} maximal. The first maximum-cardinality set to be processed is $X_3=\{2,4,5,7,8,9\}$, i.e. all sets $X_j$ contained in $X_3$ need to be determined. A moment's thought shows that the indices $j$ are exactly the elements in

$Zeros[1]\cap Zeros[3]\cap Zeros[6]= \{1,3,9,11,12,15\}$

(because $\{1,3,6\}$ is the complement of $X_3$ in $[9]$). Since all these $X_j$ (except $X_3$ itself!) are non-maximal we drop $3$ from $\{1,3,9,\ldots,15\}$ and subtract the rest:

$Candidates:=Candidates\setminus  \{1,9,11,12,15\}=\{2,3,4,5,6,7,8,10,13,14\}.$

The second maximum-cardinality set in line is $X_7=\{1,2,5,6,7,9\}$, and so upon computing

$Zeros[3]\cap Zeros[4]\cap Zeros[8]= \{1,2,7,15\}$

we put

$Candidates:=Candidates\setminus  \{1,2,15\}=\{3,4,5,6,7,8,10,13,14\}.$

The last maximum-cardinality set  is $X_{14}=\{1,3,4,6,7\}$, and so upon computing

$Zeros[2]\cap Zeros[5]\cap Zeros[8]= \{4,8,9,14\}$

we put

$Candidates:=Candidates\setminus  \{4,8,9\}=\{3,5,6,7,10,13,14\}.$

It is clear that $AllMaxsets\subseteq Candidates$. Furthermore, 
because the sets $X_3, X_7, X_{14}$ are all distinct (which we derive from $3, 7,14\in Candidates$, rather than by ad hoc testing) we
put $FoundMaxsets:=\{3,7,14\}$.  Since $Card[5]=\emptyset$ all sets (with indices) in $Card[4]=\{2,5,8,10\}$ which have not yet been deleted are maximal in ${\cal G}_0$. These are exactly the sets in $Card[4]\cap Candidates=\{5,10\}$. We hence process (at most) $X_5$ and $X_{10}$. As to $X_5=\{2,3,5,8\}$, we get

$Zeros[1]\cap Zeros[4]\cap Zeros[6]\cap Zeros[7]\cap Zeros[9]= \{5,6,10,11,12,15\}$,

and deleting that index set (except $5$ itself) yields

$Candidates:=Candidates\setminus  \{6,10,11,12,15\}=\{3,5,7,13,14\}.$

Since $10\not\in Candidates$ the set $X_{10}$  needs not be processed (it duplicates $X_5$, and generally all duplications are likewise pruned). Hence $FoundMaxsets:=\{3,7,14,5\}$.
Similar to above all sets in $Card[3]=\{1,4,6,12,13\}$ which have not yet been deleted must be maximal in
${\cal G}_0$. In view of $Card[3]\cap Candidates=\{13\}$ the only such set is $X_{13}$. Therefore $FoundMaxsets=\{3,7,14,5,13\}=Candidates$, and so
$AllMaxsets=\{3,7,14,5,13\}$.

The described method readily dualizes to sieve all {\it minimal} members of an arbitrary set family. 

Independent of maximality or minimality VL can also be used to decide whether any given set $X\subseteq U$ is $\alpha$-frequent with respect to some given database (=binary table). Namely, if $\{a,b,\ldots,c\}$ is the complement of $X$ within $U$ then $X$ is $\alpha$-frequent iff $|Zeros[a]\cap \cdots\cap Zeros[c]|\ge\alpha$.

\section{Using Vertical Layout to find all facets of any (decidable) simplicial complex}

Let $U$ be any  set (our {\it universe}). A {\it simplicial complex} is any hereditary family ${\cal S}{\cal C}$ of subsets (= {\it faces}) of $U$, i.e. $X\subseteq Y\in {\cal S}{\cal C}$ implies $X\in {\cal S}{\cal C}$.
The simplicial complexes ${\cal S}{\cal C}\subseteq {\cal P}(U)$  tackled here need to be {\it decidable} in the sense that for each $X\in  {\cal P}(U)$ one can  decide whether or not $X\in  {\cal S}{\cal C}$. The following notation will be handy:

(2)\quad $Y\hspace{-0.6mm}\hspace{-1mm}\uparrow\ :=\{X\in{\cal P}(U):\ X\supseteq Y\}$ and $Y\hspace{-0.6mm}\hspace{-1mm}\downarrow\ :=\{X\in{\cal P}(U):\ X\subseteq Y\}$

{\bf 4.1} Our algorithm, call it {\it Find-All-Facets}, starts by extending the face $\emptyset$ to larger and larger faces (by adding or discarding one random element $x\in U$ at a time) until the obtained face $F_1$ can no longer be extended. (Here we need that ${\cal S}{\cal C}$ is decidable.) Thus $F_1$ is the first found facet. If $\{a,b,\ldots, c\}$ is the complement of $F_1$ in $U$ then each face not contained in $F_1$ must contain at least one of $a,b,\ldots,c$. In other words, the face must be in the set filter
 $\{a\}\uparrow\cup\  \{b\}\uparrow\cup\cdots\cup \{c\}\uparrow$.
We can thus assume by induction that $t\ge 1$ facets $F_1,\ldots,F_t$ have been found, as well as sets $G_1,\ldots,G_s$ such that each face $X\in {\cal S}{\cal C}$ exclusively belongs to either $F_1\downarrow\cup\cdots\cup F_t\downarrow$ or
   $G_1\!\uparrow\cup\cdots\cup G_s\!\uparrow$. 

The induction step from $t$ to $t+1$ works as follows. 

{\it Case 1:} Some $G_i$ is a face of ${\cal S}{\cal C}$. Then extend $G_i$ to some facet $F_{t+1}$ (in the manner shown above). If $\{a,b,\ldots,c\}$ is the complement of $F_{t+1}$ in $U$ then each face $X$ not contained in any of the facets $F_1$ to $F_{t+1}$ (and only these $X$) must thus belong to the set filter

$\bigg( G_1\uparrow\cup \cdots \cup\ G_s\uparrow\bigg)\setminus F_{t+1}\downarrow\ \hspace{0.5cm}=$

$\bigg( (G_1\cup\{a\})\!\uparrow\cup \cdots \cup(G_1\cup\{c\})\!\uparrow\bigg)\ \cup\cdots\cup\ \bigg( (G_s\cup\{a\})\!\uparrow\cup \cdots \cup(G_s\cup\{c\})\!\uparrow\bigg)$ 

Upon relabelling the sets $G_1\cup\{a\},\ldots,G_s\cup\{c\}$ as $H_1,\ldots,H_{s'}$ we see that this proves the induction step.

{\it Case 2:} No $G_i$ is a face of ${\cal S}{\cal C}$. Then a fortiori no superset of a $G_i$ is a face, and so $F_1$ to  $F_t$ are all the facets there are.

So far Find-All-Facets does not involve Vertical Layout. But repeated application of VL is crucial from a practical point of view: Going from $G_1,\ldots,G_s$
to $H_1,\ldots, H_{s'}$ without VL would trigger  an explosion of set filter generators. That's why  VL is called to efficiently sieve the $m$
minimal sets among  $H_1,\ldots, H_{s'}$. Relabel them as
 $H_1,\ldots, H_m$. Last not least, all non-faces among $H_1,\ldots, H_m$ should be deleted. In fact, it seems better to do that first and afterwards apply VL.

{\bf 4.2}  We speculate that Find-All-Facets will prove useful  in plenty situations, not just in data mining, but let us focus on data mining. Namely, if $c: {\cal P}(E) \ra {\cal P}(E)$ is a closure operator one calls $X \subseteq E$ {\it independent} if $c(X \setminus \{x\}) \neq c(X)$ for all $x\in X$. Further, if $U$ is any closed set, then an inclusion-minimal set $Y \subseteq U$ with $c(Y) = U$ is called a {\it minimal key (for} $U)$.  It is known (though no too  well) that for all $X \subseteq E$ it holds that:

(3) \quad $X$ is independent $\Leftrightarrow X$ is a minimal key 

Several data mining applications need to know the family ${\cal S}{\cal C}$ of all minimal keys (e.g. for Association Rule  Mining). In view of (3) it is clear that ${\cal S}{\cal C}$ is a simplicial complex. Therefore Find-All-Facets (possibly followed by Facets-To-Faces) can be used.

{\bf 4.3} In some scenarios one only knows some facets $F_1,\ldots, F_t$ but  not the coupled sets $G_1,\ldots, G_s$ in 4.1. Since each face not in $F_1\downarrow\cup\cdots\cup F_t\downarrow$ must intersect all complements $U\setminus F_i$ one can calculate {\it once}\footnote{Previous methods (Dualize+Advance [GKMT] and its variants) {\it keep on} doing that. Thus for each $t$ they calculate all minimal transversals $G_i$ of the sets $U\setminus F_i\ (1\le i\le t)$. If no $G_i$ is a face then all facets have been found. Otherwise one random $G_i$ gets extended to $F_{t+1}$ while the other $G_j$'s are thrown away!  Repeated from scratch calculation of all minimal transversals of set systems  (aka hypergraph dualization) is clearly more expensive than repeatedly applying VL.}
 the sets $G_i$ as the minimal transversals of these complements, and then continue with the calculation of $F_{t+1},\ F_{t+2},\ldots$ as described in 4.1. A case in point (up to duality) is the set filter of all cutsets of a graph. At first many generators (=minimal cutsets) can be obtained fast (say the first $t$ of them) but then the process gets stuck [W3].

\section{The second toy database}

Table 4 is a database less structured than Table 1. It hence induces a simplicial complex ${\cal F}{\cal S}_2$ of frequent sets (again $\alpha=2$) whose facets are are harder to retrieve. In 5.1 we use the method of Section 4 to calculate them. Interestingly VL will be used for {\it two} unrelated purposes: a) for deciding whether sets $X\subseteq U= [9]$ are frequent and b) for finding the minimal members of various intermediate set filters. 
With the facets at hand we compress the whole of ${\cal F}{\cal S}_2$ in 5.2.

To unclutter notation we will e.g. write $24\uparrow$ for $\{2,4\}\uparrow$.

\begin{tabular}{l|c|c|c|c|c|c|c|c|c|} 
& 1 & 2& 3 & 4 & 5 &6 & 7 & 8 & 9 \\ \hline
$t_1=$ & $x$ & $x$ & $x$ &  & $x$ &  $x$& $x$ &  & $x$\\ \hline
 $t_2=$&  & $x$ &  & $x$ & $x$ &  &  & $x$ & $x$ \\ \hline 
$t_3=$ &  &  & $x$ & $x$ &  & $x$ & $x$ & $x$& $x$ \\ \hline
$t_4=$ & $x$ & $x$ & $x$ & $x$ & $x$ & $x$ & $x$ & $x$ &  \\ \hline
$t_5=$ & $x$ & $x$ &  & $x$ & $x$ &  & $x$ & $x$ & $x$ \\ \hline
$t_6=$ &$x$  &  & $x$ &  &  & $x$ & $x$ &   & $x$  \\ \hline \end{tabular}

{\sl Table 4: The seven maximal frequent sets of this database are not obvious}

{\bf 5.1} By inspection the frequent set $F_1:=13679:=\{1,3,6,7,9\}\in t_1\cap t_6$ is clearly maximal. Although at this stage a second facet $F_2$ could again be found by inspection, let us launch our systematic procedure. Since $2458$ is the complement of $F_1$ in $[9]$, we can find $F_2$ in

$2\hspace{-0.6mm}\hspace{-1mm}\uparrow\cup 4\hspace{-0.6 mm}\hspace{-1mm}\uparrow\cup 5\hspace{-1mm}\uparrow\cup 8\hspace{-1mm}\uparrow.$

All four (set filter) generators $2,\ 4,\ 5,\ 8$ happen to be frequent.  For instance $2$ extends to $F_2:=123567\in t_1\cap t_4$. Its complement being $489$ the next facet $F_3$ is to be found in

$(2\hspace{-1mm}\uparrow\cup 4\hspace{-1mm}\uparrow\cup 5\hspace{-1mm}\uparrow\cup 8\hspace{-1mm}\uparrow)\setminus F_2\hspace{-0.6mm}\hspace{-1mm}\downarrow$

$=(24\hspace{-1mm}\uparrow \cup 28\hspace{-1mm}\uparrow\cup 29\hspace{-1mm}\uparrow)\cup (4\hspace{-1mm}\uparrow \cup 48\hspace{-1mm}\uparrow\cup 49\hspace{-1mm}\uparrow)\cup
(54\hspace{-1mm}\uparrow \cup 58\hspace{-1mm}\uparrow\cup 59\hspace{-1mm}\uparrow)\cup (84\hspace{-1mm}\uparrow \cup 8\hspace{-1mm}\uparrow\cup 89\hspace{-1mm}\uparrow)=4\hspace{-1mm}\uparrow \cup 8\hspace{-1mm}\uparrow\cup 29\hspace{-1mm}\uparrow \cup 59\hspace{-1mm}\uparrow$

All four generators are frequent. We can e.g. extend $4$  to the facet $F_3:=24589\in t_2\cap t_5$. Its complement being $1367$ we conclude 

$(4\hspace{-1mm}\uparrow \cup 8\hspace{-1mm}\uparrow\cup 29\hspace{-1mm}\uparrow \cup 59\hspace{-1mm}\uparrow)\setminus 24589\hspace{-1mm}\downarrow$

$=(41\hspace{-1mm}\uparrow \cup 43\hspace{-1mm}\uparrow\cup 46\hspace{-1mm}\uparrow \cup 47\hspace{-1mm}\uparrow)\cup (81\hspace{-1mm}\uparrow \cup 83\hspace{-1mm}\uparrow\cup 86\hspace{-1mm}\uparrow \cup 87\hspace{-1mm}\uparrow)$

$\ \cup(291\hspace{-1mm}\uparrow \cup 293\hspace{-1mm}\uparrow\cup 296\hspace{-1mm}\uparrow \cup 297\hspace{-1mm}\uparrow)\cup(591\hspace{-1mm}\uparrow \cup 593\hspace{-1mm}\uparrow\cup 596\hspace{-1mm}\uparrow \cup 597\hspace{-1mm}\uparrow).$

The set $293$ (and whence each set in $293\!\uparrow$) is infrequent. Therefore $293\!\uparrow$ and like wise $296\!\uparrow,\ 593\!\uparrow,\ 596\!\uparrow$ must be deleted.
 One can e.g. extend $597$ to
 the facet $F_4:=59712\in t_1\cap t_5$. Then

$\big(41\hspace{-1mm}\uparrow \cup\cdots \cup 87\hspace{-1mm}\uparrow\cup 291\hspace{-1mm}\uparrow \cup 297\hspace{-1mm}\uparrow\cup 591\hspace{-1mm}\uparrow \cup 597\hspace{-1mm}\uparrow\big)\setminus 57912\hspace{-1mm}\downarrow$

$=\big(41\hspace{-1mm}\uparrow \cup\cdots \cup 87\hspace{-1mm}\uparrow\big)\cup 2913\hspace{-1mm}\uparrow\cup 2914\hspace{-1mm}\uparrow\cup 2916\hspace{-1mm}\uparrow\cup 2918\hspace{-1mm}\uparrow
        \cup 2973\hspace{-1mm}\uparrow\cup 2974\hspace{-1mm}\uparrow\cup 2976\hspace{-1mm}\uparrow\cup 2978\hspace{-1mm}\uparrow$
				
         $\cup 5913\hspace{-1mm}\uparrow\cup 5914\hspace{-1mm}\uparrow\cup 5916\hspace{-1mm}\uparrow\cup 5918\hspace{-1mm}\uparrow
\cup 5973\hspace{-1mm}\uparrow\cup 5974\hspace{-1mm}\uparrow\cup 5976\hspace{-1mm}\uparrow\cup 5978\hspace{-1mm}\uparrow$

$=41\hspace{-1mm}\uparrow \cup 43\hspace{-1mm}\uparrow\cup 46\hspace{-1mm}\uparrow \cup 47\hspace{-1mm}\uparrow\cup 81\hspace{-1mm}\uparrow \cup 83\hspace{-1mm}\uparrow\cup 86\hspace{-1mm}\uparrow \cup 87\hspace{-1mm}\uparrow.$

As to the last equality, note that 2+2+2+2 cancellations are due to two generators 
being contained in $41\hspace{-1mm}\uparrow$, two in $81\hspace{-1mm}\uparrow$, two in $47\hspace{-1mm}\uparrow$, two in $87\hspace{-1mm}\uparrow$ ; 
another 8 generators were infrequent. Upon extending (say)$87$ to the facet $F_5:=87346\in t_3\cap t_4$ one gets

$(41\hspace{-1mm}\uparrow \cup 43\hspace{-1mm}\uparrow\cup 46\hspace{-1mm}\uparrow \cup 47\hspace{-1mm}\uparrow\cup 81\hspace{-1mm}\uparrow \cup 83\hspace{-1mm}\uparrow\cup 86\hspace{-1mm}\uparrow \cup 87\hspace{-1mm}\uparrow)\setminus 87346\hspace{-1mm}\downarrow$

$=\cdots= 41\hspace{-1mm}\uparrow \cup 472\hspace{-1mm}\uparrow\cup 475\hspace{-1mm}\uparrow \cup 479\hspace{-1mm}\uparrow\cup 81\hspace{-1mm}\uparrow \cup 872\hspace{-1mm}\uparrow\cup 875\hspace{-1mm}\uparrow \cup 879\hspace{-1mm}\uparrow.$

Here all cancellations are due to infrequent generators, the details are left to the reader. Upon extending say $41$ to the facet $F_6:=412578\in t_4\cap t_5$ one gets

$(41\hspace{-1mm}\uparrow \cup 472\hspace{-1mm}\uparrow\cup \cdots \cup 879\hspace{-1mm}\uparrow)\setminus 412578\hspace{-1mm}\downarrow\ 
=\cdots=\ 479\hspace{-1mm}\uparrow\cup 879\hspace{-1mm}\uparrow.$

Upon extending say $879$ to the facet $G_7:=8794\in t_3\cap t_5$ one calculates

$(479\hspace{-1mm}\uparrow \cup 879\hspace{-1mm}\uparrow)\setminus 8794\hspace{-1mm}\downarrow$

$=4791\hspace{-1mm}\uparrow\cup 4792\hspace{-1mm}\uparrow\cup 4793\hspace{-1mm}\uparrow\cup 4795\hspace{-1mm}\uparrow
         \cup 4796\hspace{-1mm}\uparrow\cup 8791\hspace{-1mm}\uparrow\cup 8792\hspace{-1mm}\uparrow\cup 8793\hspace{-1mm}\uparrow
\cup 8795\hspace{-1mm}\uparrow\cup 8796\hspace{-1mm}\uparrow.$

Since all ten  generators are infrequent, we conclude that ${\cal F}{\cal S}_2=F_1\hspace{-1mm}\downarrow\cup\cdots\cup F_7\hspace{-1mm}\downarrow$.

{\bf 5.2} As in Section 2, applying the Facets-To-Faces algorithm to $F_1, \ldots, F_7$ yields $|{\cal F}{\cal S}_2|=12+42+\cdots+16=173$ frequent sets, packed in seven 012e-rows:

\begin{tabular}{l|c|c|c|c|c|c|c|c|c|c} 
& 1 & 2& 3 & 4 & 5 &6 & 7 & 8 & 9 \\ \hline
$\rho_1=$ & $2$ & $0$ & $e$ & $0$ & $0$ &  $e$& $2$ &$0$  & $1$& 12\\ \hline
 $\rho_2=$& $e_2$ & $e_2$ &$e_1$  & $0$ & $e_2$ &$e_1$  &$2$  & $0$ & $0$& 42 \\ \hline 
$\rho_3=$ & $0$ &$e_2$  & $0$ & $e_1$ &$e_2$  & $0$ & $0$ & $e_1$& $1$& 9 \\ \hline
$\rho_4=$ & $e$ & $e$ & $0$ & $0$ & $e$ & $0$ & $2$ & $0$ & $1$& 14 \\ \hline
$\rho_5=$ & $0$ & $0$ &$e$  & $2$ & $0$ & $e$ & $2$ & $2$ & $0$& 24 \\ \hline
$\rho_6=$ & $e$ & $e$ &$0$  & $2$ & $e$ & $0$ & $2$ & $2$ & $0$& 56 \\ \hline
$\rho_7=$ &$0$  &$0$  & $0$ &$2$  &$0$  & $0$ & $2$ &$2$   & $2$& 16  \\ \hline 
\end{tabular}

{\sl Table 5: Compressed representation of ${\cal F}{\cal S}_2$   }

The conclusion of this preliminary draft is as follows. While the best way to find the maximal frequent sets remains debatable and also depends on the peculiarities of the database, our method of compression seems harder to beat, particularly when the facets are large. For instance [W2] it took Facets-To-Faces 1114 seconds to compress approximately $10^{92}$ faces (contained in 70 random facets $F_i \subseteq [2000]$ each of cardinality 300) into $707'518$ many $012e$-rows. (It is known that {\it each} simplicial complex can arise as the family of all frequent sets of a suitable database.)
      
{\bf References}

\begin{enumerate}

\item[{[GKMT]}] Gunopulos, D., Khardon, R., Mannila, H., Toivonen, H.: Data mining, Hypergraph Transversals, and Machine Learning. In: Proc. of PODS 1997, pp. 209–216.

\item[{[HKMT]}] M. Holsheimer, M. Kersten, H. Mannila, H. Toivonen, A perspective on databases and data mining, KDD-95 Proceedings.

\item[{[H]}] J. Heaton, Comparing dataset characteristics that favor the Apriori, Eclat, or FP-Growth Frequent Itemset mining algorithms, South-East Conference 2016, pages 1-7.

\item[{[W1]}] M. Wild, Computations with finite closure systems and implications, LNCS 959 (1995) 111-120.

\item[{[W2]}] M. Wild, ALLSAT compressed with wildcards: Partitionings and face-numbers of simplicial complexes. Submitted.

\item[{[W3]}] M. Wild, ALLSAT compressed with wildcards: All spanning trees. In preparation

\end{enumerate}

\end{document}